\documentclass[twocolumn,10pt]{article}
\usepackage[margin=2cm]{geometry}
\usepackage{amsmath,amssymb}
\usepackage{graphicx}
\usepackage{cite}
\usepackage{hyperref}
\usepackage{url}
\usepackage{float}
\usepackage{caption}
\usepackage{times}
\usepackage{textcomp}
\usepackage{flushend}

\hypersetup{colorlinks=true,linkcolor=blue,citecolor=blue,urlcolor=blue}

\title{\textbf{Early Experiments on Macroscopic Quantum Tunneling}}
\author{Willem den Boer\\Brighton, Michigan, USA}
\date{June 2026}

\begin{document}
\maketitle

\begin{abstract}
Before conclusive evidence of Macroscopic Quantum Tunneling (MQT) was
published by Clarke, Devoret and Martinis in 1985~\cite{Devoret1985,Martinis1985}, several
other groups reported experimental results interpreted as MQT. The
first, in chronological order, was in 1980 based on studies done at
Leiden University in 1979. This paper looks back at these experiments on
low capacitance Niobium point contacts in an rf-SQUID (radio-frequency
Superconducting Quantum Interference Device) configuration at
temperatures between 1 and 4.2 K. The research was inspired by the
theoretical predictions by Ivanchenko and Zil'berman in 1969~\cite{Ivanchenko1969} on
MQT in current-biased Josephson junctions and by Leggett in 1978~\cite{Leggett1978}
on MQT in closed loops with a superconducting weak link.
\end{abstract}

\section{Introduction}

In 1985 John Clarke, Michel Devoret and John Martinis published two
seminal papers~\cite{Devoret1985,Martinis1985} which provided definite proof of macroscopic
quantum tunneling (MQT) and energy level quantization in superconducting
circuits with Josephson tunnel junctions. Their meticulous experiments
with well-defined parameters led to the Nobel prize for Physics award in
2025.

Ivanchenko and Zil'berman, theoretical physicists at the Donetsk
Physico-technical Institute in Ukraine, had first introduced the concept
already in 1969~\cite{Ivanchenko1969}, although not under the term Macroscopic Quantum
Tunneling. In section 3 of their paper, they matter-of-factly describe
quantum transitions of the phase in a current-biased Josephson junction
with the so-called tilted washboard potential:

\emph{``If \(\Theta\) (i.e. kT) = 0, then in the classical case the system will
be found in some state with a fixed phase difference and, consequently,
with V = 0. In the quantum-mechanical case the system can tunnel into a
neighboring quasi-stationary state.''}

They proceed with the quantum analog of the Hamiltonian for the system
and the probability equation of a phase slip via tunneling. Apparently,
they considered this obvious, while others, including our group at
Leiden, wondered about the philosophical implications of extending
quantum tunneling to macroscopic systems such as electrical circuits ten
million times larger than an atom. The 1985 papers from the group at the
University of California in Berkeley cited several earlier experimental
studies of Josephson junction circuits interpreted as evidence for MQT.
Two were performed on current driven Josephson junctions, respectively
on 1 \(\mu\)m\textsuperscript{2} tunnel junctions below 100 mK~\cite{Voss1981} and on
high current density junctions at 1.6 K~\cite{Jackel1981}. Two others used
magnetic flux controlled superconducting loops with point contacts~\cite{denBoer1980,Prance1981}.
In addition, other Ukrainian scientists performed pioneering,
underreported research in this field as well, recently summarized by
Turutanov~\cite{Turutanov2025}. The current paper recalls the early experiments at
Leiden University, which were, to our knowledge, the first experimental
evidence of MQT, although not conclusive.

\section{Retrospective}

During the 1970's research was ongoing at Leiden University on rf SQUIDs
using Nb point contacts. Rudolf de Bruyn Ouboter was leading the efforts
in this field. He was a professor in low temperature physics and made
significant contributions in the fields of superfluid helium, cooling
techniques, superconductivity and Josephson junctions. He was also a
resident historian and wrote several articles about the discovery of
superconductivity by Kamerlingh Onnes~\cite{deBruynOuboter1997,deBruynOuboter1987} at Leiden University.
Professor de Bruyn Ouboter passed away in February 2026 at the age of 92
after a remarkable career in experimental low temperature physics.

The author of this paper had the opportunity to join his group at the
end of 1978, having received his engineering degree several years
earlier in the dynamic group at Delft University started by professor
J.E. Mooij. The graduate work in Delft involved superconducting Sn
microbridges in tunable microwave cavities (this research did,
unfortunately, not lead to a predicted step in the current voltage curve
corresponding to the resonance frequency). I was interviewed for the
postgraduate position at the Kamerlingh Onnes Laboratory by professors
K.W. Taconis and R. de Bruyn Ouboter, not realizing at the time that
they had led the effort to experimentally demonstrate the first dilution
refrigerator allowing cooling to 220 mK~\cite{Das1965} and later to 50 mK.
This work followed de Bruyn Ouboter's doctoral dissertation in 1961:
\emph{``Thermodynamic Properties of Liquid \(^{3}\)He--\(^{4}\)He Mixtures''}.

The following is a discussion of the 1979 results at Leiden with a
caveat. As an outsider in the field of superconducting quantum
electronics for decades, I am not in a position to give a rigorous
analysis taking into account the numerous papers which led to greatly
improved understanding of macroscopic quantum phenomena. The
retrospective and analysis presented here is based on the prevailing
thought processes at the time in 1979, some of which were superseded by
later theoretical and experimental studies.

Research that occurred decades ago is not always easy to reconstruct,
but the paper we published~\cite{denBoer1980} gives details. What follows is a
summary of the experiments based on the paper, on my notes and on my
recollection. A brief analysis is also included based on our
understanding and the simplest available models at the time.

The original purpose of the project was to study flux transitions in
rf-SQUID type devices with one weak link in the superconducting loop.
Tunnel junctions based on thin film deposition and microfabrication were
not available, so the focus was on Nb point contacts. The skilled
workshop had ample experience making high quality devices, although
reproducibility was never as good as with tunnel junctions. The samples
consisted of two cylindrical blocks of Niobium separated by a thin foil
and with two finely threaded holes for Niobium screws with sharpened
points (figure 1). A cylindrical hole formed in the center of the two
blocks formed the loop inductance.

\begin{figure}[!ht]
  \centering
  \includegraphics[width=0.36\columnwidth]{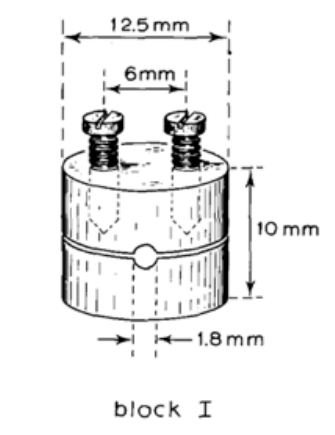}
  \hfill
  \includegraphics[width=0.56\columnwidth]{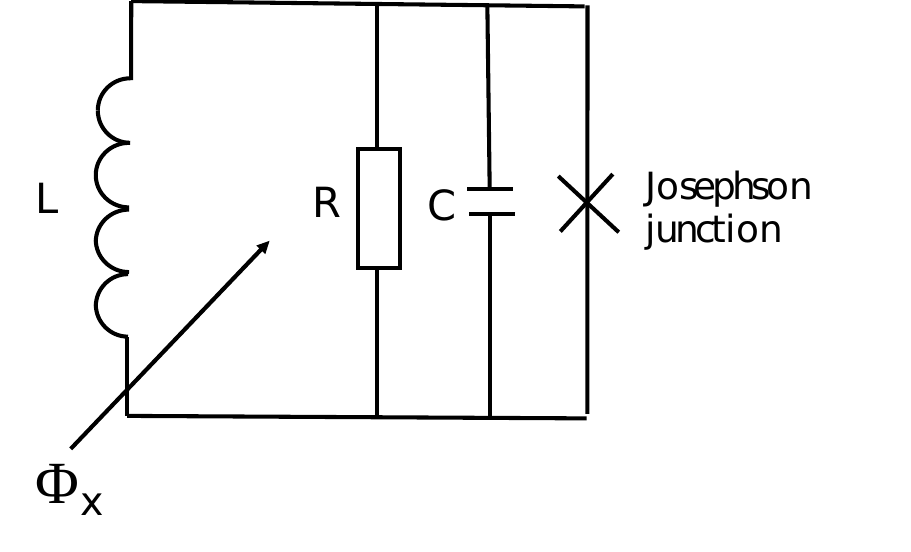}
  \caption*{\textbf{Fig1}. Sample with Nb point contacts for measurement of flux
  transitions (left)~\cite{denBoer1980} and
  equivalent circuit (right), block I refers to one of two Nb blocks used
  (reprinted with permission from Elsevier)}
  \label{fig:fig1}
\end{figure}

Both screws could be adjusted from outside the cryostat. The first one
was set to obtain a suitable critical supercurrent, while the second one
remained open. After recording the current-voltage characteristic of the
point contact, the loop was closed by short-circuiting the second point
contact. Reopening and reclosing the second point contact was used to
verify that the current-voltage of the first one had not changed. The
process of adjusting the first screw to get a critical supercurrent
\(I_{c}\) in the right range was very delicate and sometimes
frustrating. In retrospect this method seems quite primitive compared to
the current state of the art in fabricating Josephson tunnel junctions
with well-defined parameters. The author was responsible for performing
the measurements which were initially done at 4.2 K only, already well
below the critical temperature of 9.2 K for Nb.

We were inspired by the theoretical papers by Ivanchenko and Zil'berman~\cite{Ivanchenko1969}
and in particular Leggett~\cite{Leggett1978} who suggested the possibility of MQT under
the right conditions in superconducting loops with a Josephson junction.
Professor Leggett, who was awarded the Nobel prize for physics in 2003
(with V.L. Ginzburg and A.A. Abrikosov) for pioneering contributions to
the theory of superconductors and superfluids, passed away in March 2026.

His theoretical prediction of MQT in 1978 led us to reduce \emph{T} to 1
K and study the system for values of the critical Josephson supercurrent
\(I_{c}\) and inductance \(L\) where the dimensionless parameter
\(\mathcal{L} = \frac{2\pi LI_{c}}{\Phi_{0}}\) is in the range between 1
to 4 (\(\Phi_{0} = \frac{h}{2e}\) is the flux quantum). This
\(\mathcal{L}\) regime had not been studied in detail vs. temperature at
the time, since rf-SQUIDs typically had a higher value of
\(\mathcal{L}\). Dilution refrigeration was not available or likely
could not be used because of the size of the sample and the need for
adjustment from outside the cryostat, so the lowest temperature we could
achieve with \textsuperscript{4}He cooling was around 1 K.

An external magnetic field was applied via a coil and a commercial rf
SQUID was used for measuring trapped flux inside the loop via a flux
transformer. The sample was carefully shielded against external noise by
copper and lead shielding inside the cryostat, a mu-metal shield outside
the cryostat and rf filters in all leads into the cryostat. The trapped
flux was plotted vs. the applied flux on an X-Y recorder in this
\(\mathcal{L}\) regime of interest by slowly varying the current through
the external coil.

The theoretical relationship between applied and trapped flux is
well-known:

\begin{equation}
\Phi = \Phi_{x} - LI_{c}\sin\!\left(\frac{2\pi\Phi}{\Phi_{0}}\right)
\end{equation}

where \(\Phi\) is the trapped magnetic flux and \(\Phi_{x}\) the applied
magnetic flux. This relationship is shown in figure 2 for
\(\mathcal{L}\) = 4.

\begin{figure}[!ht]
  \centering
  \includegraphics[width=\columnwidth]{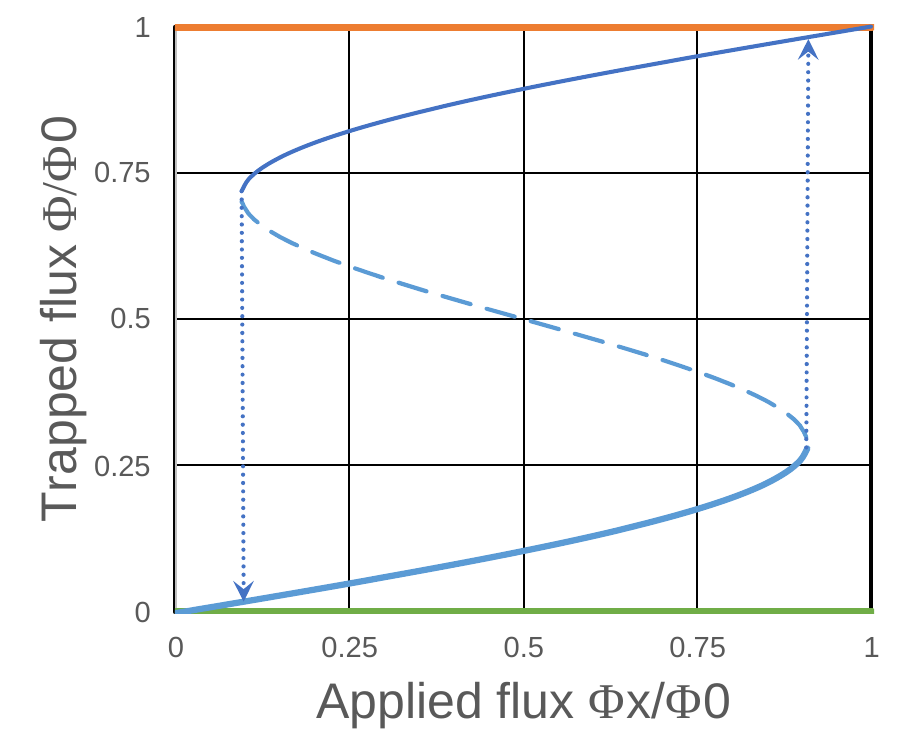}
  \caption*{\textbf{Fig2}. Theoretical dependence without noise of normalized
  trapped flux on normalized applied flux for \(I_{c} = \infty\) (fully
  closed loop, red and green lines) and
  \(I = I_{c}\sin\!\left( \frac{2\pi\Phi}{\Phi_{0}} \right)\) (blue curve
  for loop with \(\mathcal{L}\) =4, in the absence of any noise
  dotted blue lines show hysteresis)}
  \label{fig:fig2}
\end{figure}

While in an open loop the trapped flux is obviously equal to the applied
flux, in a fully closed loop without a weak link a screening
supercurrent is induced when a magnetic field is applied in the ring, so
as to keep the trapped flux quantized at multiples of the flux quantum
\emph{\(\Phi_{0}\)}. In the intermediate situation with the
Josephson junction in the loop this screening current is limited by the
critical supercurrent \(I_{c}\) of the Josephson junction. The wave
function representing Cooper pair condensates on each side of the
junction develop a phase difference \(\delta\)\emph{~= 2\(\pi\) \(\Phi\)/\(\Phi_{0}\)}.
Hence, for a screening current less than \(I_{c}\), \emph{\(\delta\)}~is
directly linked and proportional to the trapped flux \(\Phi\). In the absence
of any noise phase slips and flux jumps occur when the absolute current
increases to the critical supercurrent \(I_{c}\) of the junction
according to the Josephson equation \(I = I_{c}\sin\delta\) at \(\delta\)~=~\(\pi\)/2,
and at \(\delta\)~=~\(-\pi\)/2~in figure 2. This results in hysteresis.

The potential energy \(U\) vs. trapped flux is given by:

\begin{equation}
U = - \frac{\Phi_{0}}{2\pi}I_{c}\cos\!\left( \frac{2\pi\Phi}{\Phi_{0}} \right) + \ \frac{1}{2L}{(\Phi - \Phi_{x})}^{2}
\end{equation}

A straightforward numerical calculation shows that at an applied flux
\(\Phi_{x} = 0.5\,\Phi_{0}\)~~metastable minima occur for a
representative set of baseline parameters from one of the experiments
(\emph{L} = 1.1 nH, \(I_{c}\) = 1.2 \(\mu\)A corresponding to
\(\mathcal{L}\) = 4). They are separated by a relatively
shallow potential barrier (figure 3). Small variations in applied flux
\(\Phi_{x}\) will raise one minimum and lower the other.

\begin{figure}[!ht]
  \centering
  \includegraphics[width=\columnwidth]{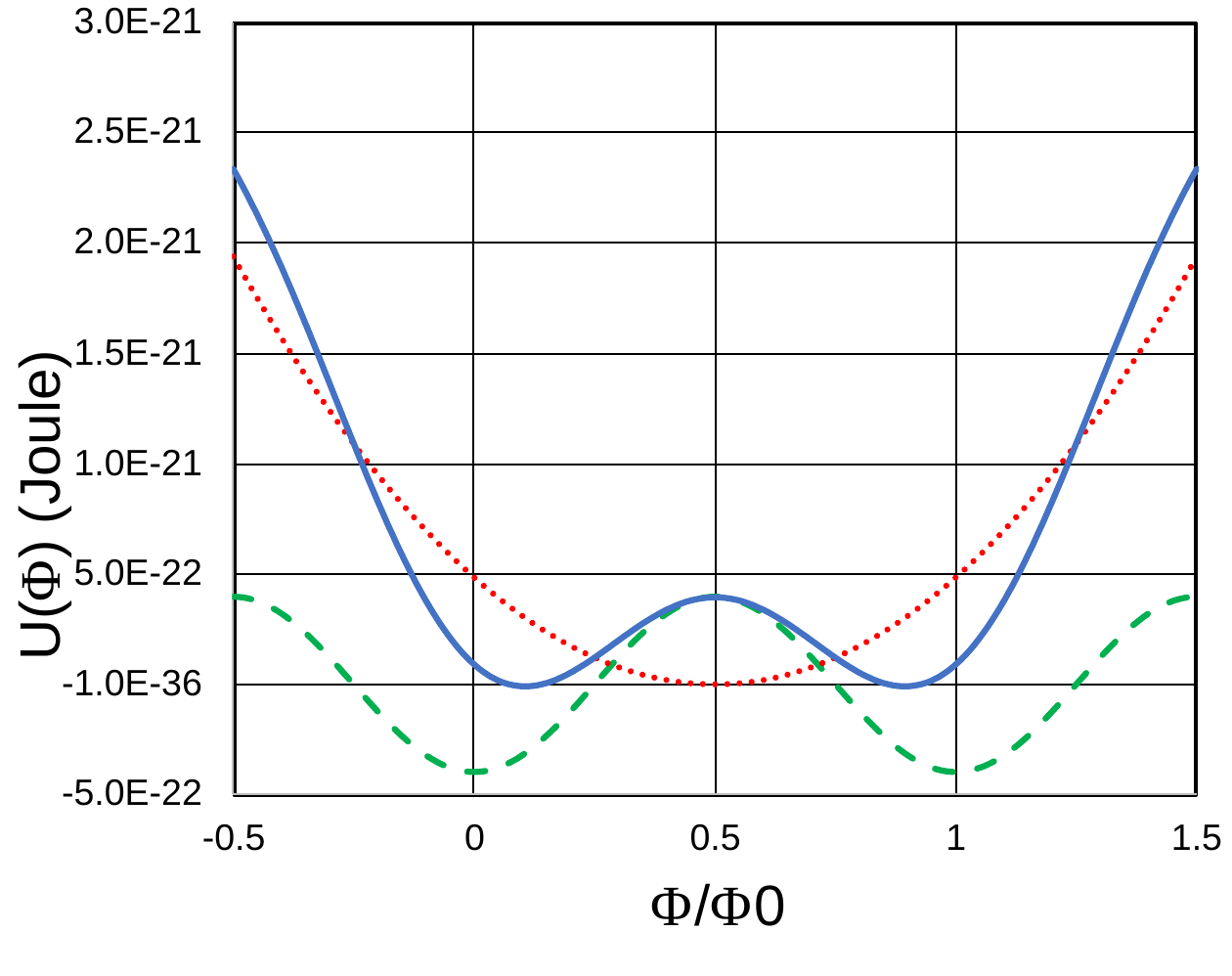}
  \caption*{\textbf{Fig3}. Potential energy of the ring system vs. trapped flux for
  \(\mathcal{L}\) = 4 at \(\Phi_{x} = 0.5\Phi_{0}\), solid blue curve is U
  for L = 1.1 nH and \(I_{c}\)= 1.2 \(\mu\)A, dashed green curve is cos term
  in equation 2, dotted red curve is quadratic term}
  \label{fig:fig3}
\end{figure}

Leggett~\cite{Leggett1978} noted the analogy of this system with a particle having
mass attempting to cross a potential barrier via tunneling in
geometrical space and suggested MQT was possible with \emph{\(\Phi\)} replacing
the positional coordinate x and the capacitance \emph{C} replacing mass
\emph{m}.

The fictitious particle with ``mass \emph{C}'' can escape across the
potential barrier between the two metastable states by thermal
excitation or by tunneling (MQT) or by a combination of the two. Both
will reduce or eliminate the hysteresis in figure 2. MQT, based on
Leggett's paper, was hypothetical at the time of the experiments. In the
simplest model without dissipation the escape rates for thermal
excitation \(R_{\text{th}}\) over and for tunneling \(R_{\text{tun}}\)
through the barrier are, respectively:

\begin{equation}
R_{\text{th}} = \frac{\omega_{a}}{2\pi}\exp\!\left(\frac{- \Delta U}{\text{kT}}\right)
\end{equation}

and

\begin{equation}
R_{\text{tun}} = \frac{\omega_{a}}{2\pi}\exp\!\left\{ \frac{- 2}{\hbar}\int_{A}^{B}{\sqrt{2C(U\left( \Phi \right) - U_{0})}\,d\Phi} \right\}
\end{equation}

where \emph{\(\omega_{a}\)/2\(\pi\)}~~is the attempt to
escape frequency, \emph{\(\Delta U = U_{\text{max}} -
U_{\text{min}}\)} is the barrier height and
\emph{\(U_{0}\)} is the minimum energy in the well plus the
zero point energy (\emph{\(U_{\text{min}}\)} +
\(\hbar\omega_{a}/2\)). A and B are the beginning and endpoint for
tunneling through the barrier.

The equation for tunneling is based on the WKB
(Wentzel-Kramers-Brillouin) method for calculating tunneling of single
particles with mass m through an arbitrarily shaped barrier. To
calculate the escape rates in equations 3 and 4 we used measured values
of \emph{L} and \(I_{c}\) and published or upper estimates of values for
\emph{C}. \emph{R} was derived from the asymptotic values of the
current-voltage curve (300 ohm in the baseline experiment mentioned
above), but does not appear in the simplest model without dissipation.

The total inductance \emph{L} is the sum of the loop conductance and the
point contact inductance and can be derived from the slope of the flux
curve at \(\Phi_{x} = 0\) using equation (1), when \(I_{c}\) is known.
The current-voltage curves of the point contact, measured using a
current source, for values of \(I_{c}\) in the range of interest showed
no hysteresis, implying an upper limit for the capacitance of
\(C = \frac{\hbar}{2eI_{c}R^{2}}\) (3 fF for \emph{R} = 300 ohm and
\(I_{c} = 1.2\,\mu\)A).

We could not measure \emph{C} of the point contacts directly. It was
known at the time that for sharp point contacts the capacitance was very
small (0.1 to 1 fF) based on its geometry and the observation of very
high plasma frequencies up to 200 GHz in point contacts~\cite{Tolner1975}. In
tunnel junctions \emph{C} is much larger and better defined. Equation 4
shows that a very small capacitance favors tunneling over thermal
escape. For the calculations it was assumed to be 1 fF.

It is clear from equations 3 and 4 that the thermal escape rate is to a
large degree independent of \emph{C} and the tunneling rate is
independent of temperature, as shown in figure 4 for the baseline
parameters listed above and compatible with the experiment.

\begin{figure}[t]
  \centering
  \includegraphics[width=0.92\columnwidth]{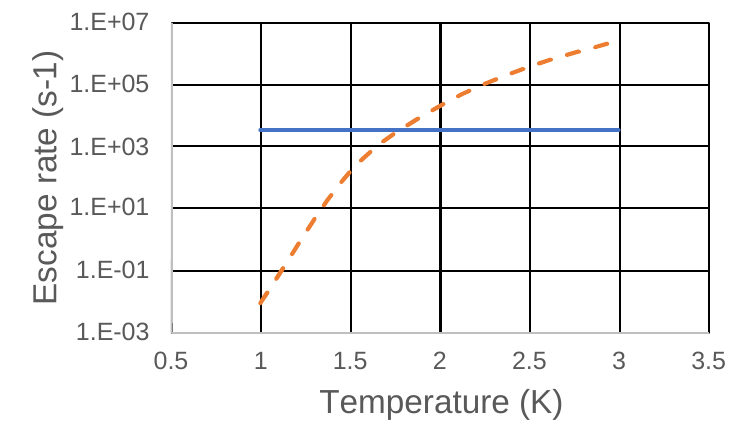}\\[1pt]
  \emph{(a)}\\[3pt]
  \includegraphics[width=0.92\columnwidth]{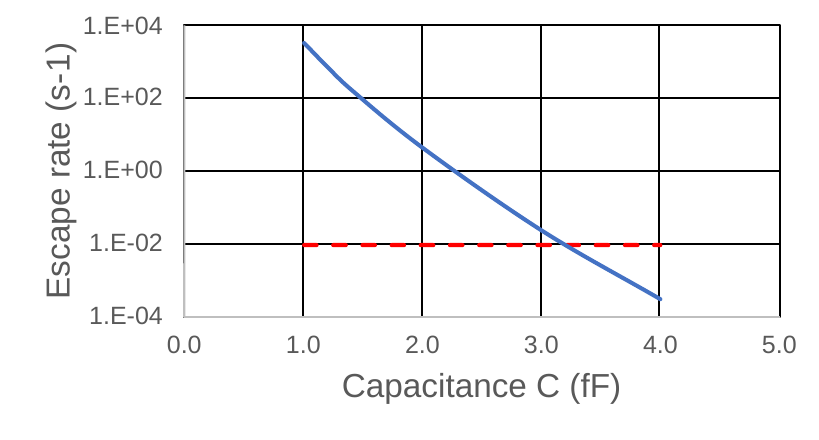}\\[1pt]
  \emph{(b)}
  \caption*{\textbf{Fig4}. Calculated dependence of \emph{R\textsubscript{th}} (red,
  dashed curves) and \emph{R\textsubscript{tun}} (blue, solid curves) on T
  (a) and on C at 1 K (b) in the absence of dissipation when the other
  baseline parameters are fixed}
  \label{fig:fig4}
\end{figure}

In the case of the current driven circuit without loop the attempt to
escape frequency from the potential wells is the plasma frequency of the
Josephson junction. In the SQUID type system it is more likely the
resonance frequency of the LC circuit, (\emph{LC})\textsuperscript{-0.5}
in the case of low damping or \emph{R/L} in the case of high damping of
our experiments. When calculating the values of plasma and resonance
frequencies they are within a range of ten. Their exact value (which is
reduced for shallow potential wells) is in a first approximation not
important for the crossover temperature between thermal escape and
tunneling, because the attempt to escape frequency is the same for
tunneling and thermal escape. Hence, the ratio of \(R_{\text{th}}\) to
\(R_{\text{tun}}\) is independent of \emph{\(\omega_{a}\)} (from
equation 3 and 4). The temperature where tunneling and thermal escape
rates are equal, \(T_{\text{crossover}}\), is, in the simplest model,
given by:

\vspace{-2pt}
\begin{equation}
\frac{1}{T_{\text{crossover}}} = \frac{- 2k}{\hbar\Delta U}\int_{A}^{B}{\sqrt{2C(U\left( \Phi \right) - U_{0})}\,d\Phi}
\end{equation}

In figure 4(a) \(T_{\text{crossover}}\) is 1.7 K, below which
tunneling could dominate in the absence of dissipation and other noise
sources.

A large number of flux curves were measured for different settings of
the first point contact in the range \(\mathcal{L}\) = 1 to 4 at 4.2 K
and 1 K. Escape rates could be observed on an oscilloscope~\cite{denBoer1980} for a
range of frequencies compatible with the specifications of the
commercial SQUID. A couple of representative flux curves at 4.2 K and 1
K~\cite{denBoer1980} are shown in figure 5. Curves at 4.2 K were all continuous
without hysteresis and showed a strong dependence of the escape rates on
10\% variations in \emph{T}, as observed on the oscilloscope. The
calculated variation in \(R_{\text{th}}\) for a 10\% variation is a
factor of about 75 for the barrier height in the experiment.
Measurements were in qualitative agreement and thermal escape was
therefore clearly dominant at 4.2 K.

A curve at 1 K with \(I_{c}\) = 1.2 \(\mu\)A, L= 1.1 nH, R= 300 ohm and C = 1
fF (a reasonable upper limit for point contacts), corresponding to the
baseline parameters of the calculations, is shown in figure 5 as well.
The calculated \(R_{\text{th}}\) is low, less than 0.01
s\textsuperscript{-1} at this temperature, 4 10\textsuperscript{5} times
lower than \(R_{\text{tun}}\) (see figure 4(a)), and would, in the
absence of tunneling, cause significant hysteresis in the flux curve. No
hysteresis was observed down to 1 K. When changing the temperature by 10\%
for this curve, the measured escape rate did not significantly
change, while thermal excitations without tunneling would again lead to
a factor of about 75. This weak temperature dependence of escape rates
in SQUIDs was also observed by Dmitrenko et al.~\cite{Dmitrenko1982} at 0.5 K.

\begin{figure}[!ht]
  \centering
  \includegraphics[width=\columnwidth]{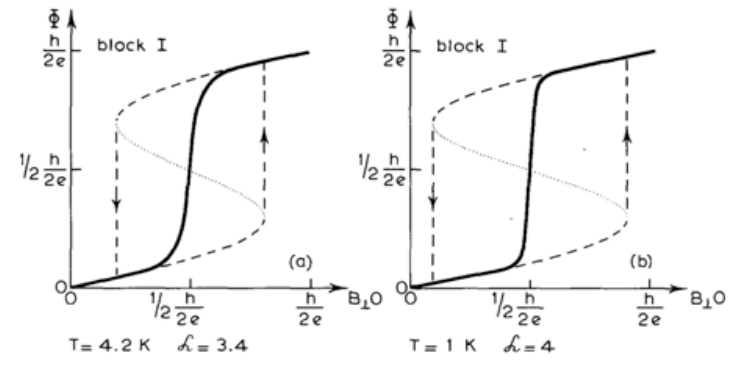}
  \caption*{\textbf{Fig5}. Representative flux curves measured at 4.2 K and at 1 K,
  where \(B_{\perp}O\)~~is the applied flux and block I refers to
  figure 1~\cite{denBoer1980} (reprinted with permission from Elsevier)}
  \label{fig:fig5}
\end{figure}

Based on the calculations of the escape rates, the absence of hysteresis
in the flux curve and the very weak dependence of the observed escape
rates on 10\% temperature variations we came to the conclusion that MQT
likely caused the flux transitions at 1 K.

\section{Brief analysis}

In 1982 Caldeira and Leggett~\cite{Caldeira1981} published their paper on the
effect of dissipation on tunneling in general and on MQT specifically.
They calculated that the tunneling rate is reduced by a factor of up to
\(\exp\!\left\lfloor - \frac{A\left( \Delta\Phi \right)^{2}}{\hbar R} \right\rfloor\)
where A is about 1, \(\Delta\Phi\) is the distance under the barrier and
\emph{R} is the resistance parallel to the junction in figure 1. This
calculation was for tunneling from a metastable minimum through a
potential barrier into a lower minimum and would reduce
\(R_{\text{tun}}\) by several orders of magnitude in the presence of
dissipation mostly caused by quasi-particle currents. It meant that the
condition \(\hbar\omega_{a} \gg kT\) needed to be satisfied for
tunneling to dominate. In our point contact with \emph{C} = 1 fF and
\(I_{c}\) = 1.2 \(\mu\)A the ratio of
\(\hbar\omega_{a}\) to \(kT\) at 1 K is calculated to be
14.5 if the Josephson plasma frequency is used for \(\omega_{a}\). It is
lower if (\emph{LC})\textsuperscript{-0.5} or \emph{R/L} is assumed.

It has been argued recently that the asymptotic \emph{R} value derived
from the current-voltage curve may be subject to a UV cutoff frequency~\cite{Altimiras2025},
which may alter the tunneling rate in the above equation.
Murani et al.~\cite{Murani2020} interpreted experiments on SQUIDS as the absence
of a dissipative quantum phase transition in resistively shunted
Josephson junctions. A theory supporting this observation was presented
by Altimiras et al.~\cite{Altimiras2025}, but has not yet led to a consensus in the
scientific community.

It is interesting to note that there are significant differences between
the escape mechanisms for a current-biased Josephson junction and for a
closed loop with a Josephson junction with applied flux of
\(0.5\Phi_{0}\). In the current-biased Josephson junction the phase
difference \(\delta\) is the quantum variable and the potential energy vs. \(\delta\)~~has
multiple wells in the shape of the tilted washboard. Escape over the
barrier results in the fictitious particle rolling down the washboard
and a DC voltage across the junction.

In the closed loop, on the other hand, the quantum variable is the
trapped flux in the loop and there are only two potential wells in the
\(\mathcal{L}\) region of interest. When the ``particle with mass
\emph{C}'' traverses the barrier, it is trapped in the second well and
can only escape again by returning to the first well. This leads, around
\(\Phi_{x} = 0.5\Phi_{0}\), to reversible phase shifts and alternating
clockwise and counterclockwise supercurrents. The current varies between
positive and negative subcritical supercurrents resulting in a small AC
voltage according to \(V = \frac{\hbar\dot{\delta}}{2e}\),
while in the current driven junction there is a DC current just below
\(I_{c}\) with continuous phase slips and resulting larger DC voltage.
It could be argued that dissipation is therefore less for the ring
configuration as compared to the current biased Josephson junction. The
ring may operate closer to a dissipation free system as pointed out by
Turutanov~\cite{Turutanov2025}, since the Josephson junction remains superconducting
with only a tiny AC voltage with amplitude below 1 picoV appearing
across the junction. On the other hand the ring system is very sensitive
to flux and photonic noise. Even if dissipation in the ring system
reduces the tunneling rate by four orders of magnitude, tunneling will
remain larger than thermal escape at 1K, as shown in figure 4(a).

The escape rates can also be calculated by varying \emph{L} or \(I_{c}\)
while keeping all other parameters at the baseline values. This is shown
in figure 6. At 1 K the \(R_{\text{tun}}\) is higher than
\(R_{\text{th}}\) over the range of values simulated. The dependence of
critical current \(I_{c}\) on magnetic field was not directly
measured in our experiments up to \(\Phi_{x} = \Phi_{0}\), but did not
vary significantly in this range, since the slopes of the flux curve at
\(\Phi_{x} = 0\) and \(\Phi_{x} = \Phi_{0}\) were the same.

\begin{figure}[t]
  \centering
  \includegraphics[width=0.92\columnwidth]{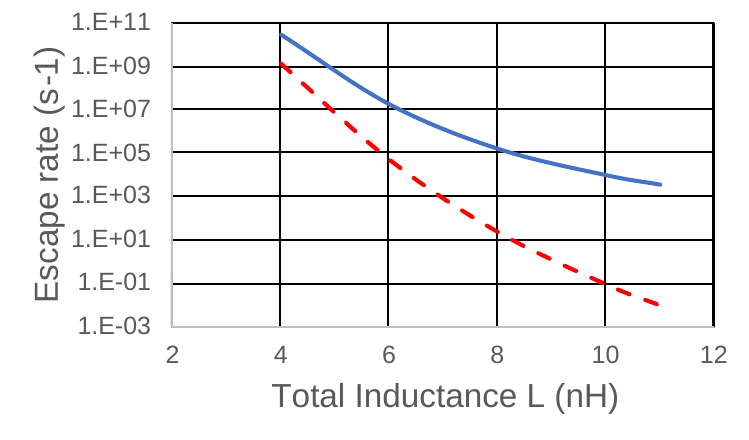}\\[1pt]
  \emph{(a)}\\[3pt]
  \includegraphics[width=0.92\columnwidth]{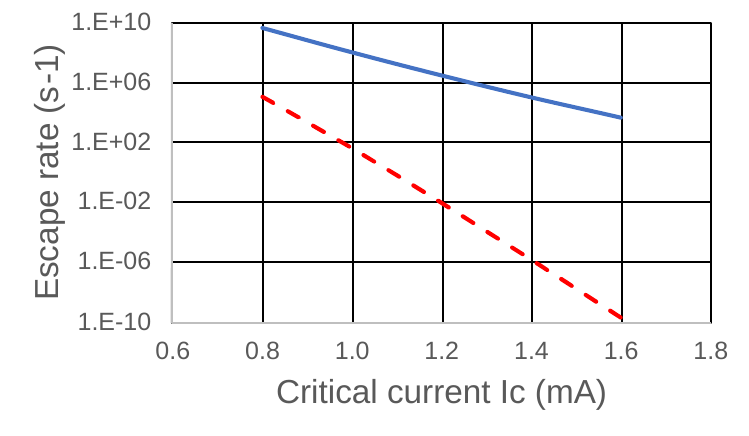}\\[1pt]
  \emph{(b)}
  \caption*{\textbf{Fig6}. Calculated dependence of \emph{R\textsubscript{th}} (red,
  dashed lines) and \emph{R\textsubscript{tun}} (blue, solid lines) at 1 K
  on total inductance of the ring and on critical current, again keeping
  all other parameters constant at the baseline values}
  \label{fig:fig6}
\end{figure}

\noindent Major noise sources are thermal excitations (Johnson noise) and magnetic
flux noise. Low frequency flux noise may be simulated by varying the
externally applied flux \(\Phi_{x}\) by small amounts around 0.5
\(\Phi_{0}\), as shown in figure 7. This makes the potential wells
asymmetric. In the simple model the ratio between \(R_{\text{tun}}\) and
\(R_{\text{th}}\) does not change much.

Compared to tunnel junctions there was uncertainty in the parameters of
our circuit, specifically the capacitance and the non-ideal Josephson
behavior of point contacts. Possible noise sources other than thermal
and flux noise may have played a role as well, in particular broadband
photonic noise from radiation leaking into the sample from the
environment. It seemed difficult, however, to explain the experimental
results without the phenomenon of MQT. One of the main differences
between tunnel junctions and point contacts is the ultralow capacitance
of point contacts which made MQT or a combination of MQT and thermal
escape likely at 1 K without the need to further cool the sample.

\section{Perspective}

Tunneling of individual particles, each with its own wavefunction, e.g.
in alpha decay and in diodes and other semiconductor devices, was
well-known long before MQT experiments. Electron tunneling has been
widely commercialized and is ubiquitous, e.g. in the flash memory used
in every smart phone and USB memory stick. The fact that the phenomenon
could be extrapolated to the quantum variables \(\delta\) and \(\Phi\) in macroscopic
systems such as superconducting electrical circuits was not obvious in
the 1970's and early 1980's, except for Ivanchenko and Zil'berman. It
underscored the validity of quantum mechanics on a wider scope and
eventually led to the design of electrical circuits behaving like
artificial atoms.

\begin{figure}[!ht]
  \centering
  \includegraphics[width=\columnwidth]{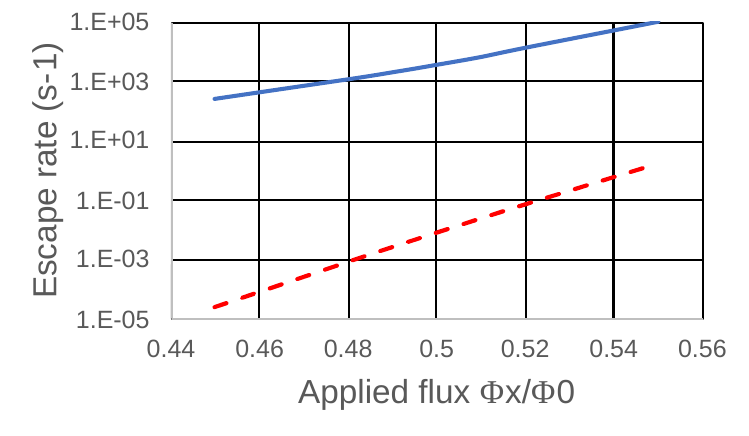}
  \caption*{\textbf{Fig7}. Calculated dependence of rates for
  \emph{R\textsubscript{th}} (red, dashed line) and
  \emph{R\textsubscript{tun}} (blue, solid line) at 1 K on applied flux
  variation around 0.5 \(\Phi_{0}\)}
  \label{fig:fig7}
\end{figure}

It is nearly impossible to reconstruct the early Leiden experiments
after so many decades. The original draft of our paper listed in the
abstract the phrase ``MQT plays a role, which becomes dominant at lower
temperatures''. In the final submission for publication professor de
Bruyn Ouboter added a caveat and changed this phrase to ``MQT might play
a role, which becomes dominant at lower temperatures'', probably
realizing that some of the assumptions on dissipation and noise made a
firm conclusion not yet possible. The UC Berkeley group of Clarke,
Devoret and Martinis ended all speculation with conclusive evidence of
not only MQT, but also energy quantization in the potential wells of
current-biased Josephson junctions~\cite{Clarke1988}. After the discovery of MQT
it took almost fifteen years to demonstrate superconducting qubits with
Macroscopic Quantum Coherence~\cite{Nakamura1999,Mooij1999} which accelerated the field of
Quantum Computing.

Interwell coherence led to fluxon type qubits. Intrawell coherence based
on energy quantization led to transmon type qubits.

The author of this paper ended up in industry working on large are
electronics on glass after receiving a doctoral degree in Electrical
Engineering at Delft University of Technology on amorphous silicon
devices, a much more mundane topic with ample commercial applications
nevertheless. It is not well-known that amorphous silicon thin film
transistors, invented at the University of Dundee in 1979, are still
used in the vast majority of LCD TVs and monitors~\cite{denBoer2005}. The reader
of this article may well be watching millions of these transistors in
action while viewing this paper online. The expectation is that
superconducting qubits will have an impact on computing that exceeds the
impact of amorphous silicon on display technology.

\section*{Acknowledgements}

The author thanks professor Carlo Beenakker of Leiden University for his
support in submitting this article. I am also grateful to professor
Sergey Frolov of the University of Pittsburgh for reading a draft of
this article and helpful suggestions.

\bibliographystyle{unsrt}
\bibliography{bibliography}

@article{Devoret1985,
  author  = {Devoret, M.H. and Martinis, J.M. and Clarke, J.},
  title   = {Measurement of Macroscopic Quantum Tunneling out of a Zero-Voltage State of a Current-Biased Josephson Junction},
  journal = {Phys. Rev. Lett.},
  volume  = {55},
  pages   = {1908--1911},
  year    = {1985}
}

@article{Martinis1985,
  author  = {Martinis, J.M. and Devoret, M.H. and Clarke, J.},
  title   = {Energy-level quantization in the zero-voltage state of a current-biased Josephson junction},
  journal = {Phys. Rev. Lett.},
  volume  = {55},
  pages   = {1543--1546},
  year    = {1985},
  doi     = {10.1103/PhysRevLett.55.1543}
}

@article{Ivanchenko1969,
  author  = {Ivanchenko, Y.M. and Zil'berman, L.A.},
  title   = {The Josephson effect in small tunnel contacts},
  journal = {Sov. Phys. JETP},
  volume  = {28},
  number  = {6},
  pages   = {1272--1276},
  year    = {1969}
}

@article{Leggett1978,
  author  = {Leggett, A.J.},
  title   = {Prospects in ultralow temperature physics},
  journal = {J. Phys. Colloques},
  volume  = {39},
  number  = {C6},
  pages   = {1264--1269},
  year    = {1978},
  doi     = {10.1051/jphyscol:19786555}
}

@article{Voss1981,
  author  = {Voss, R.F. and Webb, R.A.},
  title   = {Macroscopic quantum tunneling in 1-$\mu$m Nb Josephson junctions},
  journal = {Phys. Rev. Lett.},
  volume  = {47},
  pages   = {265--268},
  year    = {1981},
  doi     = {10.1103/PhysRevLett.47.265}
}

@article{Jackel1981,
  author  = {Jackel, L.D. and Gordon, J.P. and Hu, E.L. and Howard, R.E. and Fetter, L.A. and Tennant, D.M. and Epworth, R.W. and Kurkijarvi, J.},
  title   = {Decay of the zero-voltage state in small-area, high-current-density Josephson junctions},
  journal = {Phys. Rev. Lett.},
  volume  = {47},
  pages   = {697--700},
  year    = {1981},
  doi     = {10.1103/PhysRevLett.47.697}
}

@article{denBoer1980,
  author  = {den Boer, W. and de Bruyn Ouboter, R.},
  title   = {Flux transition mechanisms in superconducting loops closed with a low capacitance point contact},
  journal = {Physica B+C},
  volume  = {98},
  number  = {3},
  pages   = {185--191},
  year    = {1980},
  doi     = {10.1016/0378-4363(80)90075-3}
}

@article{Prance1981,
  author  = {Prance, R.J. and Long, A.P. and Clarke, T.D. and Widom, A. and Mutton, J.E. and Sacco, J. and Potts, M.W. and Megaloudis, G. and Goodall, F.},
  title   = {Macroscopic quantum electrodynamic effects in a superconducting ring containing a Josephson weak link},
  journal = {Nature},
  volume  = {289},
  pages   = {543--549},
  year    = {1981}
}

@article{Turutanov2025,
  author  = {Turutanov, O.G.},
  title   = {Nobel Prize and the contribution of Ukrainian scientists to the understanding of quantum phenomena, in particular the behavior of macroscopic quantum systems (Nobel Prize in Physics 2025)},
  journal = {Visn. Nac. Akad. Nauk Ukr.},
  volume  = {12},
  pages   = {20--30},
  year    = {2025},
  doi     = {10.15407/visn2025.12.020}
}

@article{deBruynOuboter1997,
  author  = {de Bruyn Ouboter, R.},
  title   = {Heike Kamerlingh Onnes's Discovery of Superconductivity},
  journal = {Scientific American},
  volume  = {276},
  number  = {3},
  pages   = {98--103},
  year    = {1997}
}

@article{deBruynOuboter1987,
  author  = {de Bruyn Ouboter, R.},
  title   = {Superconductivity: Discoveries during the Early Years of Low Temperature Research at Leiden 1908--1914},
  journal = {IEEE Trans. Magn.},
  volume  = {23},
  number  = {2},
  pages   = {355--370},
  year    = {1987}
}

@inproceedings{Das1965,
  author    = {Das, T.P. and De Bruyn Ouboter, R. and Taconis, K.W.},
  title     = {in Proc. 9th Int. Conf. on Low Temp. Phys.},
  booktitle = {Proc. 9th Int. Conf. on Low Temp. Phys., Columbus, Ohio},
  editor    = {Daunt, J.G. and Edwards, D.O. and Milford, F.J. and Yaqub, M.},
  publisher = {Plenum Press},
  address   = {New York},
  pages     = {1253},
  year      = {1965}
}

@article{Tolner1975,
  author  = {Tolner, H. and Andriesse, C.D.},
  title   = {High impedance point contact Josephson junctions},
  journal = {IEEE Trans. Magn.},
  volume  = {11},
  pages   = {866--870},
  year    = {1975}
}

@article{Dmitrenko1982,
  author  = {Dmitrenko, I.M. and Tsoi, G.M. and Shnyrkov, V.I.},
  title   = {Macroscopic quantum tunneling in a system with dissipation},
  journal = {Sov. J. Low Temp. Phys.},
  volume  = {8},
  number  = {6},
  pages   = {330},
  year    = {1982},
  doi     = {10.1063/10.0030726}
}

@article{Caldeira1981,
  author  = {Caldeira, A.O. and Leggett, A.J.},
  title   = {Influence of dissipation on quantum tunneling in macroscopic systems},
  journal = {Phys. Rev. Lett.},
  volume  = {46},
  pages   = {211--214},
  year    = {1981},
  doi     = {10.1103/PhysRevLett.46.211}
}

@misc{Altimiras2025,
  author       = {Altimiras, C. and Esteve, D. and Girit, C. and le Sueur, H. and Joyez, P.},
  title        = {Absence of a dissipative quantum phase transition in Josephson junctions: Theory},
  howpublished = {arXiv:2312.14754v5 [cond-mat.supr-com] 15 Jan 2025},
  year         = {2025}
}

@article{Murani2020,
  author  = {Murani, A. and Bourlet, N. and le Sueur, H. and Portier, F. and Altimiras, C. and Esteve, D. and Grabert, H. and Stockburger, J. and Ankerhold, J. and Joyez, P.},
  title   = {Absence of a Dissipative Quantum Phase Transition in Josephson Junctions},
  journal = {Phys. Rev. X},
  volume  = {10},
  number  = {2},
  pages   = {021003},
  year    = {2020},
  doi     = {10.1103/PhysRevX.10.021003}
}

@article{Clarke1988,
  author  = {Clarke, J. and Cleland, A.N. and Devoret, M.H. and Esteve, D. and Martinis, J.M.},
  title   = {Quantum Mechanics of a Macroscopic Variable: The Phase Difference of a Josephson Junction},
  journal = {Science},
  volume  = {239},
  pages   = {992--997},
  year    = {1988}
}

@article{Nakamura1999,
  author  = {Nakamura, Y. and Pashkin, Yu.A. and Tsai, J.S.},
  title   = {Coherent control of macroscopic quantum states in a single-Cooper-pair box},
  journal = {Nature},
  volume  = {398},
  pages   = {786--788},
  year    = {1999},
  doi     = {10.1038/19718}
}

@article{Mooij1999,
  author  = {Mooij, J.E. and Orlando, T.P. and Levitov, L. and Tian, L. and van der Wal, C.H. and Lloyd, S.},
  title   = {Josephson persistent-current qubit},
  journal = {Science},
  volume  = {285},
  pages   = {1036--1039},
  year    = {1999}
}

@book{denBoer2005,
  author    = {den Boer, W.},
  title     = {Active Matrix Liquid Crystal Displays: Fundamentals and Applications},
  publisher = {Newnes, a division of Elsevier},
  address   = {Oxford},
  year      = {2005}
}

\end{document}